# Virtual Manipulation in an Immersive Virtual Environment: Simulation of Virtual Assembly

**Mojtaba Noghabaei, SM.ASCE, [1] Khashayar Asadi, SM.ASCE, [2] and Kevin Han, Ph.D., M.ASCE [3]**

[1]Department of Civil, Construction, and Environmental Engineering, North Carolina State University, P.O. Box 7908, Raleigh, NC 27695-7908; e-mail: snoghab@ncsu.edu

[2]Department of Civil, Construction, and Environmental Engineering, North Carolina State University, P.O. Box 7908, Raleigh, NC 27695-7908; e-mail: kasadib@ncsu.edu

[3]Department of Civil, Construction, and Environmental Engineering, North Carolina State University, P.O. Box 7908, Raleigh, NC 27695-7908; e-mail: kevin_han@ncsu.edu

**ABSTRACT**

To fill the lack of research efforts in virtual assembly of modules and training, this paper presents a virtual manipulation of building objects in an Immersive Virtual Environment (IVE). A worker wearing a Virtual Reality (VR) head-mounted device (HMD) virtually perform an assembly of multiple modules while identifying any issues. Hand motions of the worker are tracked by a motion sensor mounted on the HMD. The worker can be graded based on his/her overall performance and speed during this VR simulation. The developed VR simulation can ultimately enable workers to identify unforeseen issues (e.g., not enough clearance for an object to be installed). The presented method can solve current deficiencies in discrepancy detection in 3D scanned models of elements. The developed VR platform can also be used for interactive training and simulation sessions that can potentially improve efficiency and help achieve better work performance for assemblies of complex systems.

**INTRODUCTION**

Industrial construction is one of the major parts of the construction industry in all of the countries especially industrially developing countries (Nahangi and Haas 2016). Industrial construction mainly consists of power plants, petrochemical, and manufacturing facilities (Nahangi and Haas 2016). According to the US Census Bureau, the total construction spending on power plant construction alone was more than $99 billion for both private and public entities in 2017 (US Census Bureau 2018). This sector alone was more than eight percent of the total construction that has occurred in the US in 2017 (US Census Bureau 2018).

      Piping is one of the most important systems in industrial construction, which includes piping design, fabrication, and installation. The piping systems account for nearly 50% of the total industrial construction costs (Nahangi and Haas 2016). Despite advancements in Building Information Modeling (BIM) and as-built modeling for complex piping systems, errors and discrepancies are still a problem in industrial construction. The discrepancy in piping systems is still a common problem in most of the power plants with complex piping systems. Construction engineers need to detect these discrepancies before moving the components to the construction site from manufacturing facilities so they can reduce cost and schedule overruns. However, this is a

cumbersome task since checking the components using old module surveying techniques is a labor-intensive task and requires years of experience.

To solve these problems, researchers used 3D imaging technologies along with BIM models to generate as-built 3D models and detect discrepancies (Asadi et al. 2019b; Bosché et al. 2015; Han et al. 2018; Nahangi and Haas 2016). Comparing as-built 3D models with BIM elements can help the engineers and technicians to detect the discrepancies easily(Boroujeni and Han 2017). To automatically detect the discrepancies in the elements after 3D scanning, researchers focused on micro-level point cloud analysis and noise cancellation techniques (Nahangi et al. 2016). Although this approach is well tuned for pipes, it will not work efficiently on complex piping systems where there are more complex components than pipes. This limitation of micro-level point cloud analysis on the 3D models slowed the adaptation of these technologies in a real construction site.

In this paper, the authors suggest a macro-level analysis of 3D scanned as-built models for detecting manufacturing discrepancies as wells a platform for training workers in an immersive virtual environment (IVE) that enables workers and designers to identify potential problems and discrepancies of as-built elements before transferring them to the construction site. This platform can help the workers to detect the potential problems that might happen in the construction site such as insufficient space for performing a task and the positions that might be hard to access. Moreover, this platform enables workers and designers to virtually manipulate objects using hand motion tracker and head-mounted device (HMD) in an IVE and automatically detect whether the element has passed the quality standards. Finally, the 3D scanned object snaps in its registered highlighted area in the BIM model if the object passes the quality standards. Therefore, this platform can reduce cost and time associated with quality control industrial constructions. Our contribution in this study is proposing a novel VR training platform for improving workers performance for complex assembly tasks in industrial construction.

**BACKGROUND**

This section is investigating different viewports to determine the potential gaps for detecting discrepancies in the construction industry. In the beginning, the existing tools for surveying and visualizing industrial modules and parts were investigated and both advantages and disadvantages were reviewed. Furthermore, current approaches for assessing the quality of as-built models versus BIM elements in terms of geometry in civil infrastructures were evaluated. Finally, the potentials of Virtual Reality (VR) and Virtual Manipulation technologies was discussed and the benefits of using these technologies were evaluated.

Surveying techniques are grouped based on automation level that they can provide for data gathering. The most famous technique in surveying is using conventional tapes. Since this technique was performed manually, it can be error-prone and unreliable. Moreover, in complex systems, manual measurements are labor-intensive and error-prone. Alternatively, new automated data collection techniques such as laser scanning and photogrammetry, have the potentials to decrease error and automate surveying process (Fathi et al. 2015). These techniques made substantial improvements in the project schedule, project monitoring, and as-built modeling (Asadi et al. 2018b, 2019a; Asadi and Han 2018; Golparvar-Fard et al. 2011).

Using the accurate and reliable data form the automated data collection techniques, the as-built status of each component in a construction site can be generated (Asadi et al. 2018a; Son et al. 2015). However, aligning the point cloud of each element to its twin in BIM to detect discrepancies is a manual task and yet to be automated (Nahangi and Haas 2016). Automating this task improves discrepancy detection.

Despite all the benefits of 3D scanning and BIM in detecting the discrepancies, still there are other technologies such as Virtual Reality (VR) and Virtual Manipulation (VM) that can help to improve discrepancy detection process. In a survey, the industry professionals indicated that using VR along with BIM is one of the main solutions for the future of their company and the industry (Salama et al. 2018). VR can provide a one to one scale level of BIM (Heydarian et al. 2015), therefore it is possible to improve discrepancy detection in such environments. In addition, many studies have shown, VR can improve spatial perception (Paes et al. 2017) and improving spatial perception along with one to one scale level of VR can help technicians to detect the discrepancies faster and more efficient. Furthermore, VR can improve communication between different parties (Balali et al. 2018), improve safety (State and State 2017), and provide effective training programs (Kayhani et al. 2019). In addition, combining VR with hand motion tracking technologies such as Leap Motion can provide a realistic first-person situation where the user can interact, move, and rotate the objects using their hands (Hilfert and König 2016). This technology further improves the spatial perception of the user which finally leads to easier discrepancy detection and improving user performance in detecting manufacturing defects.

**METHOD**

The proposed method consists of two main components. The first component is to generate a point cloud for each building element (i.e., a pipe) and comparing this as-built point cloud with its BIM model, finding the corresponding BIM element and finding geometrical and/or spatial differences at the macro level. The second component is an IVE platform that enables users to manipulate 3D scanned as-built models of the elements for detecting discrepancies in the parts in a one to one scale virtual environment. The authors used Unity 3D as the main platform for comparing different BIM models and 3D point clouds

In this paper, the author used three parts from a piping section and generated the laser scanned models of those pipes. Figure 1. Shows the laser scanned models of the parts with label "a" and the image of the parts by label "b".

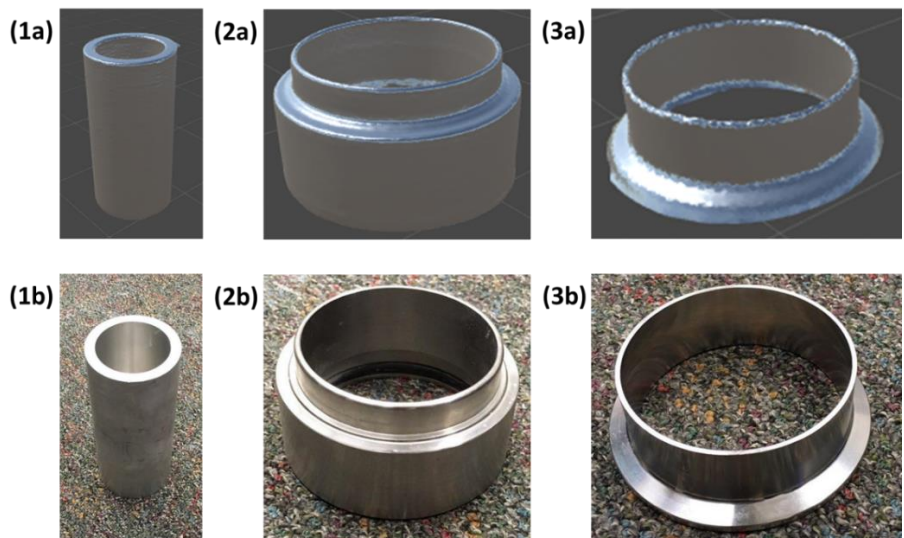

**Figure 1. Pictures vs. the 3D scanned models of the parts.**

To make 3D scanned models of the parts, the authors used the Artec Eva laser scanner (Artec Eva 2018). This hand-held scanner can achieve the accuracy up to 0.1 millimeters. This scanner enables a technician to simply walk on the construction site and/or manufacturing site and easily scan different parts. Figure 2. Represents the process of 3D scanning a pipe using the scanner. The pipe is on the rotary table while 3D hand-held 3D scanner stays fixed to generate an accurate and near noise-free 3D model. After scanning the parts, Artec Studio's automated process was used to generate a 3D mesh using the point cloud that was generated by the device.

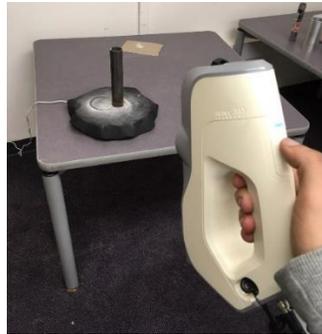

**Figure 2. The process of creating 3D models using Artec laser scanner**

To automatically compare the as-built models and BIM models, the authors defined three macro-level parameters with a certain threshold. The reason behind using macro-level parameters is the noise in 3D scanned models. 3D scanning techniques always have noise and it is the inherent deficiency of the current technologies. Both photogrammetry and laser scanning do not provide perfect and noise-less point clouds. Therefore, using micro-level parameters to compare two models can be complicated on complex elements and needs post-processing of the point cloud which can be labor-intensive and error-prone. However, using macro-level parameters can automate the alignment process for more complex elements. Table 1. defines three macro-level parameters to detect differences in the 3D scanned model and BIM model. In this table, each triangle is representing each triangle in the 3D model's mesh. $D_{max}$ and $D_{min}$ represent the maximum and minimum value of all vertices in the 3D model's mesh in direction D. $\vec{N_D}$ is the absolute value of the normalized vector of each triangles of the 3D model's mesh in direction D.

**Table 1. Macro-level parameters for detecting discrepancies in 3D scanned Models.**

| Parameter | Formula |
|---|---|
| Total Object Surface | $\sum_{each\ triangle} Area\ of\ each\ triangle$ |
| Object Dimension | $D_{max} - D_{min}$ |
| Aggregated Normals | $\dfrac{\sum_{each\ triangle} \vec{N_D} * Area\ of\ each\ triangle}{\sum_{each\ triangle} Area\ of\ each\ triangle}$ |

After defining the macro-level parameters, the authors created an immersive manipulation environment. Leap Motion that detects a user's hands is attached to an HTC Vive VR headset. This integrated hardware enables users to interact with virtual objects in an IVE using their hands. Figure 3 presents an overview of this platform. First, BIM models and laser scanned models are imported in Unity 3D. Then, connect both VR device and hand tracker to the Unity 3D. In the developed IVE, users can move, rotate, and connect virtual elements.

Using this platform, a user can place 3D point clouds in the corresponding and highlighted BIM element. If the macro-level parameters for both 3D scanned model and BIM model are within the threshold, the object will snap into the highlighted area and the quality of the part is acceptable. Otherwise, the part will not snap and it has more discrepancy than acceptable range. This means that the part has to be changed and does not satisfy the quality standards. Therefore, the user can try virtual installation and inspection of building parts in the IVE.

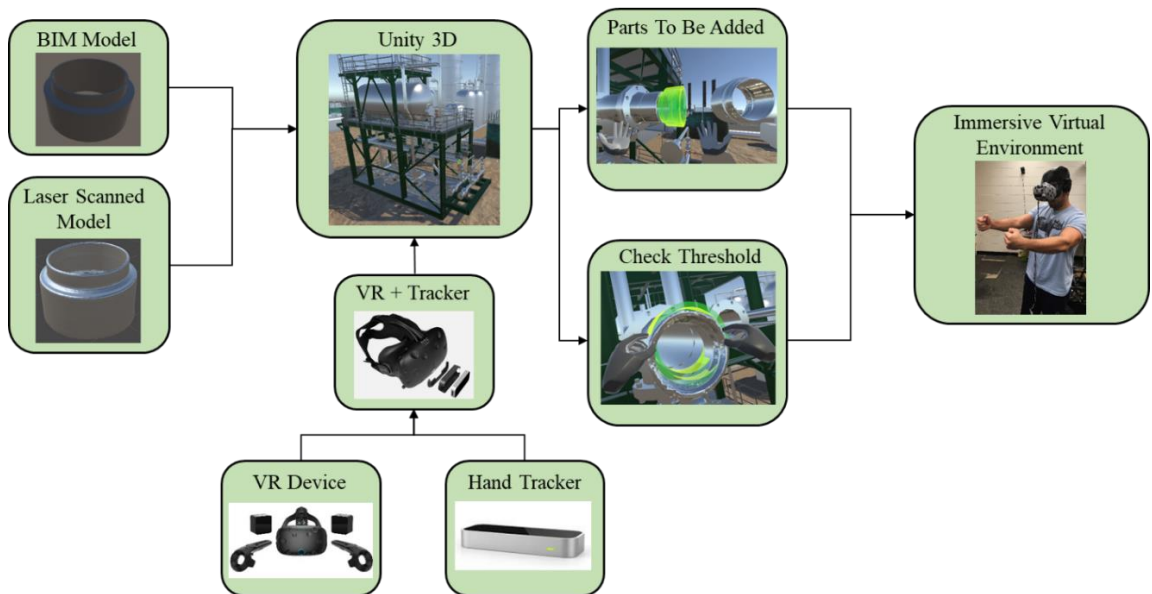

**Figure 3. Overview of the developed VR manipulation platform**

**EXPERIMENTAL SETUP AND RESULTS**

The collected data from three objects in Figure 1. are as follows. Object one has 198,402 vertices and 396,800 faces. Object two has 125,742 vertices and 251,484 faces. Object three has 19,658 vertices and 38,703 faces. To validate the platform, the authors compared the macro-level parameters values from 3D scanned models and BIM models. Table 2. represents the different macro-level parameters values for Object 2 in Figure 1. The BIM model row in Table 2. shows the macro-level parameters for BIM model and the next row shows the value of the macro-level parameter for 3D scanned objects. The difference of values between BIM and 3D scanned model are within a certain threshold. The acceptable threshold for this study assumed to be 25 percent, however, it can be changed based on the needed accuracy level and importance of the part.

Table 2. Vision-based 3D models vs. the actual 3D scanned models of the parts.

| Parameter Direction (unit) | Total Surface ($mm^2$) | Weighted Normals | | | Object Dimension | | |
|---|---|---|---|---|---|---|---|
| | | X | Y | Z | X (mm) | Y (mm) | Z (mm) |
| BIM Model | 28404.4 | 0.52 | 0.53 | 0.18 | 90.0 | 89.5 | 47.0 |
| Laser Scanned Model | 27911.9 | 0.57 | 0.57 | 0.24 | 90.9 | 92.3 | 55.1 |

Figure 4. demonstrate the first person view of the user trying to place a 3D scanned model in the highlighted area in a petrochemical plant site. In Figure 4a, the 3D scanned element is hovering in the air in the right side of the screen and the position of the element in the BIM model is highlighted by green color in the center of the screen. In Figure 4b, the user grasped the 3D scanned model. The user goal is to place the 3D scanned part in the highlighted area and make sure that the part has acceptable quality. In Figure 4c, the user moves the part close to the highlighted area and in Figure 4d, the 3D scanned model snaps in the highlighted area since it is within a certain threshold. If the object does not snap in the highlighted area, the macro-parameters for that specific object is not within the defined threshold therefore, the user has to check the part manually and detect discrepancies in part so they can fix the issue with the real part.

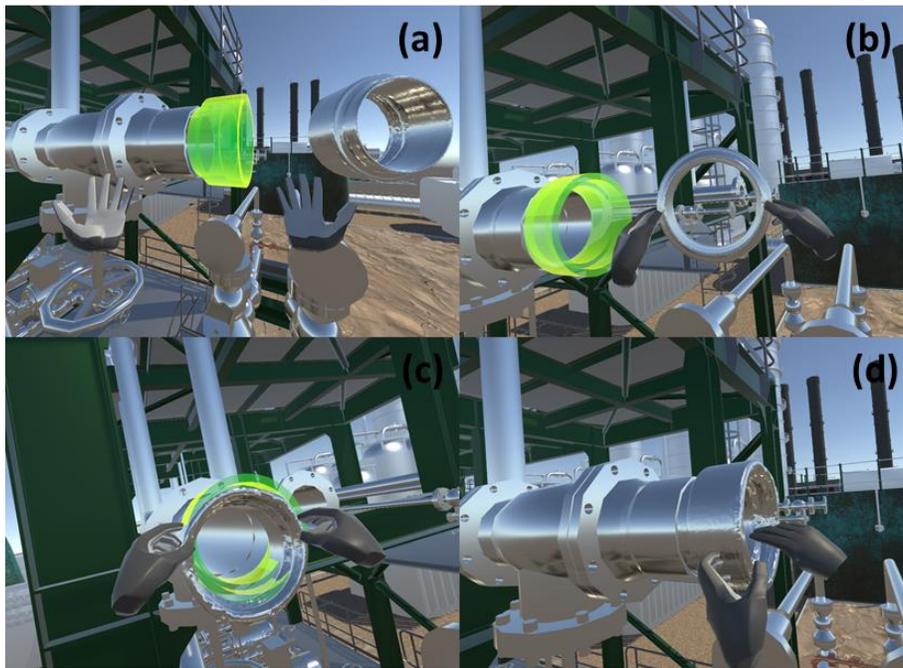

**Figure 4. Worker view in a virtual petrochemical plant working with as-built 3D parts.**

**CONCLUSION**

This study proposed a use case of virtual manipulation that can be potentially used for installation and inspection for industrial construction. The developed platform has the potential to reduce and solve on-site problems and in the module production plants and reduce unnecessary transportation of the elements. Also, this platform can help the workers to automatically detect discrepancies and avoid quality control on the parts that have high quality. In addition, the workers will perform the tasks faster in the construction site since they had proper training with virtual elements and this can help to avoid schedule overruns that might happen in a construction site.

This platform can help to check the discrepancies in parts that were produced in different companies before transferring them to the construction site. The VM environment helps

Proceedings Paper Formatting Instructions – 6 – Rev. 01/2019

technicians to check the parts and virtually assemble the modules. In addition, this platform has the potential to be used as a training program environment where technicians and workers can practice connecting complex piping systems in a safe environment.

**REFERENCES**

"Artec Eva." (2018). <https://www.artec3d.com/portable-3d-scanners/artec-eva> (Feb. 12, 2018).

Asadi, K., and Han, K. (2018). "Real-Time Image-to-BIM Registration Using Perspective Alignment for Automated Construction Monitoring." *Construction Research Congress 2018*, American Society of Civil Engineers, Reston, VA, 388–397.

Asadi, K., Jain, R., Qin, Z., Sun, M., Noghabaei, M., Cole, J., Han, K., and Lobaton, E. (2019a). "Vision-based Obstacle Removal System for Autonomous Ground Vehicles Using a Robotic Arm." Computing in Civil Engineering 2019.

Asadi, K., Ramshankar, H., Noghabaee, M., and Han, K. (2019b). "Real-time Image Localization and Registration with BIM Using Perspective Alignment for Indoor Monitoring of Construction." *Journal of Computing in Civil Engineering*.

Asadi, K., Ramshankar, H., Pullagurla, H., Bhandare, A., Shanbhag, S., Mehta, P., Kundu, S., Han, K., Lobaton, E., and Wu, T. (2018a). "Building an Integrated Mobile Robotic System for Real-Time Applications in Construction." *arXiv preprint arXiv:1803.01745*.

Asadi, K., Ramshankar, H., Pullagurla, H., Bhandare, A., Shanbhag, S., Mehta, P., Kundu, S., Han, K., Lobaton, E., and Wu, T. (2018b). "Vision-based integrated mobile robotic system for real-time applications in construction." *Automation in Construction*, Elsevier, 96, 470–482.

Balali, V., Noghabaei, M., Heydarian, A., and Han, K. (2018). "Improved Stakeholder Communication and Visualizations: Real-Time Interaction and Cost Estimation within Immersive Virtual Environments." *Construction Research Congress 2018*, 522–530.

Boroujeni, K. A., and Han, K. (2017). "Perspective-Based Image-to-BIM Alignment for Automated Visual Data Collection and Construction Performance Monitoring." *Congress on Computing in Civil Engineering, Proceedings*, 171–178.

Bosché, F., Ahmed, M., Turkan, Y., Haas, C. T., and Haas, R. (2015). "The value of integrating Scan-to-BIM and Scan-vs-BIM techniques for construction monitoring using laser scanning and BIM: The case of cylindrical MEP components." *Automation in Construction*, Elsevier, 49, 201–213.

Fathi, H., Dai, F., and Lourakis, M. (2015). "Automated as-built 3D reconstruction of civil infrastructure using computer vision: Achievements, opportunities, and challenges." *Advanced Engineering Informatics*, Elsevier, 29(2), 149–161.

Golparvar-Fard, M., Bohn, J., Teizer, J., Savarese, S., and Peña-Mora, F. (2011). "Evaluation of image-based modeling and laser scanning accuracy for emerging automated performance monitoring techniques." *Automation in Construction*, Elsevier, 20(8), 1143–1155.

Han, K., Degol, J., and Golparvar-Fard, M. (2018). "Geometry- and Appearance-Based Reasoning of Construction Progress Monitoring." *Journal of Construction Engineering and Management*, 144(2), 04017110.

Heydarian, A., Carneiro, J. P., Gerber, D., and Becerik-gerber, B. (2015). "Immersive virtual environments , understanding the impact of design features and occupant choice upon lighting for building performance." *Building and Environment*, Elsevier Ltd, 89, 217–228.

Hilfert, T., and König, M. (2016). "Low-cost virtual reality environment for engineering and construction." *Visualization in Engineering*, Springer International Publishing, 4(1), 2.

Kayhani, N., Taghaddos, H., Noghabaee, M., Ulrich, and Hermann. (2019). "Utilization of Virtual Reality Visualizations on Heavy Mobile Crane Planning for Modular Construction." *arXiv e-prints*, arXiv:1901.06248.

Nahangi, M., Czerniawski, T., Haas, C. T., Walbridge, S., and West, J. (2016). "Parallel Systems and Structural Frames Realignment Planning and Actuation Strategy." *Journal of Computing in Civil Engineering*, 30(4), 04015067.

Nahangi, M., and Haas, C. T. (2016). "Skeleton-based discrepancy feedback for automated realignment of industrial assemblies." *Automation in Construction*, Elsevier, 61, 147–161.

Paes, D., Arantes, E., and Irizarry, J. (2017). "Immersive environment for improving the understanding of architectural 3D models: Comparing user spatial perception between immersive and traditional virtual reality systems." *Automation in Construction*, Elsevier, 84, 292–303.

Salama, T., Moselhi, O., and Al-Hussein, M. (2018). "Modular Industry Characteristics and Barrier to its Increased Market Share." (April).

Son, H., Bosché, F., and Kim, C. (2015). "As-built data acquisition and its use in production monitoring and automated layout of civil infrastructure: A survey." *Advanced Engineering Informatics*, Elsevier, 29(2), 172–183.

State, N. C., and State, N. C. (2017). "Development of Immersive Personalized Training Environment for Construction Workers Idris Jeelani 1 ; Kevin Han 2 ; and Alex Albert 3 1." 407–415.

United States Census Bureau. (2018). *Monthly Construction Spending*.